\documentclass[prl,aps,twocolumn,10pt,showpacs]{revtex4-1}
\usepackage{graphicx}
\usepackage{amsmath}

\begin{document}

\title{
%Comment on
%`Randomness versus specifics for word-frequency distributions'
Dependence of exponents on text length
versus finite-size scaling
for word-frequency distributions
}

\author{\'Alvaro Corral$^{1,2,3,4}$}
\author{Francesc Font-Clos$^{5}$}

\affiliation{
$^{1}$Centre de Recerca Matem\`atica,
Edifici C, Campus Bellaterra,
E-08193 Barcelona, Spain\\
$^{2}$Departament de Matem\`atiques,
Facultat de Ci\`encies,
Universitat Aut\`onoma de Barcelona,
E-08193 Barcelona, Spain\\
$^{3}$Barcelona Graduate School of Mathematics, 
Edifici C, Campus Bellaterra,
E-08193 Barcelona, Spain\\
$^{4}$Complexity Science Hub Vienna,
Josefst\"adter Stra$\beta$e 39,
1080 Vienna,
Austria\\
$^{5}$ISI Foundation, 
%Via Alassio 11/c, 
Via Chisola 5,
10126 Torino, Italy\\
}
%\date{\today}

\begin{abstract} 
%Yan and Minnhagen [\emph{Physica A} 444, 828 (2016)] 

Some authors have recently 
argued that a finite-size scaling law for the text-length dependence of word-frequency distributions 
%proposed by Font-Clos et al. [\emph{New J. Phys.} 15, 093033 (2013)] 
cannot be conceptually valid. %and ``is not borne out by the data''.
Here we give solid quantitative evidence for the validity of such scaling law, 
both using 
careful statistical tests and
analytical arguments based on the generalized central-limit theorem
applied to the moments of the distribution
(and obtaining a novel derivation of Heaps' law as a by-product).
%the reason that such claims are mostly unjustified, with the scaling law being valid
%with excellent accuracy. 
%%for absolute frequencies greater than about 10. 
We also find that 
%the very counter-example provided by Yan and Minnhagen fulfils Font-Clos et al.'s scaling hypothesis
%with excellent accuracy, at odds with Yan and Minnhagen's criticisms.
%%, and supporting our approach. 
%Yan and Minnhagen's [\emph{Physica A} 444, 828 (2016)] 
the picture of word-frequency distributions with power-law exponents
that decrease with text length 
[Yan and Minnhagen, \emph{Physica A} 444, 828 (2016)] 
does not stand with rigorous statistical analysis.
Instead, we show that the distributions are perfectly described by power-law tails
with stable exponents, whose values are close to 2, in agreement with the classical Zipf's law.
Some misconceptions about scaling are also clarified.
\end{abstract}

\maketitle

\section{Introduction}
Many complex processes in biology, social science, economy, Internet science,
or cognitive science
are mimicked by the occurrence of words in texts.
Indeed, the statistics of 
insects in plants \cite{Pueyo}, 
molecules in cells \cite{Furusawa2003}, 
inhabitants in cities \cite{Malevergne_Sornette_umpu}, 
followers of religions \cite{Clauset}, 
telephone calls to people \cite{Clauset}, 
employees in companies \cite{Axtell}, 
links to sites in the worldwide web \cite{Adamic_Huberman},
chords in musical pieces \cite{Serra_scirep},
etc., 
share with word-frequency distributions 
the property of being broadly distributed, 
or ``heavy tailed''.
And most of these phenomena are described, at least asymptotically,
by power-law distributions with exponents close to 2;
in such cases one can talk about the fulfillment of Zipf's law \cite{Corral_Boleda,Moreno_Sanchez}.

A fundamental problem is how these systems evolve, in particular, 
how they grow to reach a state for which a power law, 
or even Zipf's law, holds 
\cite{Newman_05,Corominas_dice,Loreto_urn}.
In Ref. \cite{Minnhagen2009}, Bernhardsson, da Rocha, and Minnhagen
challenge the ``Zipf's view'',
proposing that the distribution of word 
frequencies in a text or collection of texts (of the same author) changes
with text length as
\begin{equation}
D_L(k) = A \frac{e^{-k/(c_0 L)}}{k^{\gamma(L)}},
\label{eq:scalingsuecos}
\end{equation}
where 
$k$ is the absolute frequency (number of tokens)
of the different words (word types),
$L$ is text length in number of tokens,
%%($M$ in Ref. \cite{Minnhagen2009}'s notation),
$D_L(k)$ is the probability mass function of $k$
(i.e., the distribution of word frequencies), 
$\gamma(L)$ is a power-law exponent, 
%$c_0$ is a scale parameter (independent on $L$) for $k/L$,
$c_0$ is a constant parameter (independent on $L$),
and $A$ is a normalizing constant.
Bernhardsson et al.'s equation [Eq. (\ref{eq:scalingsuecos})]
should apply to individual texts or collections 
when one considers parts of length $L$ of the whole.
The key ingredient of that approach
to model the change of $D_L(k)$ with $L$
is the explicit dependence of the exponent $\gamma$
on text length $L$, decreasing with increasing $L$.
%from $\gamma=2$ to $\gamma=1$, roughly.
Note also that Eq. (\ref{eq:scalingsuecos})
%\cite{Minnhagen2009} 
implies that, for the largest $k$, the word-frequency distribution
decays exponentially (in contrast to the algebraic decay proposed in Zipf's law).

Alternatively, in Ref. \cite{Font-Clos2013}, 
we (together with Boleda) 
%a third author) 
argue that 
the variability of the statistics of words in a text with its
length is more simply explained by a scaling law,
\begin{equation}
\label{eq:scalingfontclos}
D_L(k) = \frac 1 {L V_L} g(k/L),
\end{equation}
where $V_L$ is the size of vocabulary
(number of different words, i.e., word types)
for a fraction of text of length $L$,
and $g(z)$ is a undefined scaling function, 
the same for any value of $L$
(but not necessarily the same for different authors).

{Note that the scaling-law paradigm,
Ref. \cite{Font-Clos2013} and
Eq. (\ref{eq:scalingfontclos}), does not assume
any particular, parametric shape of $D_L(k)$ 
[in contrast to Ref. \cite{Minnhagen2009} and Eq. (\ref{eq:scalingsuecos}),
which give a truncated gamma distribution];
the scaling-law paradigm only states that, 
for a fixed text, all the $D_L(k)$'s have the same shape no matter the 
value of $L$, but at different characteristic scales given by $L$.
In other words, the shape parameters of the distributions do not change 
with $L$, 
whatever the form of this distribution is
(we do not enter here into such debate 
\cite{Ferrer2000a,Montemurro01,Li2010a,Moreno_Sanchez}),
and it is only a scale parameter what changes, proportionally to $L$.
Both exponential tails and power-law tails are allowed by the scaling function $g(z)$;
what is ``forbidden'' are text-length dependent exponents $\gamma(L)$.
Moreover, the scaling paradigm represented by Eq. (\ref{eq:scalingfontclos})
does not involve any free parameter, 
as there is no restriction on the scaling function $g$, 
and $L$ and $V_L$ are given directly by the text.
In fact, the scaling law is just a finite-size scaling law
\cite{Brezin,Barber,Privman,Corral_garciamillan}
(which was not explicitly mentioned in Ref. \cite{Font-Clos2013}).
}

Subsequently, Yan and Minnhagen claimed that this scaling law
is ``fundamentally impossible'' \cite{Yan_comment},
%and ``fundamentally incorrect''.
``cannot be conceptually valid'' and ``is not borne out by the data'' \cite{Yan_Minnhagen}.
These statements
constitute good examples of the sometimes counterintuitive nature of scaling laws.
Let us summarize the points of these authors to make it clear
that their critique is not relevant.
%essentially irrelevant.
%%as we explained previously \cite{FontClos_Corral_reply}.
\begin{itemize}
\item
First, in their Fig. 1, they find that the scaling law does not hold 
for $k=1$.
\item
Second, in their Fig. 2 they show that the scaling law does not work well for,
let us say, $k \le 10$.
\item
Third, it is argued that a ``Randomness view'', based
in the concepts of 
``Random Group Formation'', ``Random Book Transformation'', and 
``Metabook'' predicts the right form of $D_L(k)$,
which is that of Refs. \cite{Minnhagen2009,Baek2011},
i.e., Eq. (\ref{eq:scalingsuecos}) above.
\end{itemize}

It is quite clear that the first and second criticisms of Yan and Minnhagen \cite{Yan_Minnhagen}
are not fundamental, 
as they simply imply that the scaling law can only be
valid beyond the low-frequency limit, so, 
the scaling law could be rewritten as
$$
D_L(k) = \frac 1 {L V_L} g(k/L), \mbox{ for } k > 10.
$$
This is not surprising at all, as it is well known in statistical physics
that scaling laws 
are usually observed
%hold 
only asymptotically (see Appendix I at the end).
It is remarkable then that, for texts, scaling is attained just after the first
decade in frequencies (i.e., for words that appear more than about 10 times).
It is also remarkable that, despite the fact that Yan and Minnhagen 
\cite{Yan_Minnhagen}
stretch the scaling hypothesis up to very small fragments of texts
%($212473/500 \simeq 400$ tokens, for the case of \emph{Moby-Dick}),
% ^ de donde habia salido este numero??
%($215525/500 \simeq 400$ 
($215525$ tokens divided into $500$ fragments yielding about $400$ 
tokens in each one, for the case of \emph{Moby-Dick}),
the scaling law still is fulfilled reasonably well
beyond the first decade in $k$, as we detail below.
Naturally, the appropriate way to further test the validity of the scaling
law is in the opposite way, analyzing larger and larger texts.
The third point of these authors \cite{Yan_Minnhagen} is also not justified, 
as the authors do not provide any statistical evidence supporting 
the claim that Eq. (\ref{eq:scalingsuecos}) fits the empirical data
to an acceptable confidence level.

In this paper we revise the evidence for the
finite-size scaling law for word-frequency distributions,
Ref. \cite{Font-Clos2013} and Eq. (\ref{eq:scalingfontclos}),
comparing this approach with the one of
Ref. \cite{Minnhagen2009} and Eq. (\ref{eq:scalingsuecos}).
In Sec. 2 we summarize the main claims in Ref. \cite{Minnhagen2009}
and in which way they relate to the validity of the scaling law;
next, we compare the performance of different fits related to
the two approaches; 
subsequently we use a direct method to test the validity of the scaling hypothesis  
applied to word-frequency distributions; 
and then we compare with other ways of assessing errors in scaling.
The penultimate section presents a novel theoretical calculation 
of the scaling of the moments of the distribution using the generalized central-limit theorem, 
connecting them with the scaling law 
(and yielding a derivation of Heaps' law as a by-product). 
The conclusions and two appendices are at the end.
As the empirical evidence in favor of the scaling law
used in Ref. \cite{Font-Clos2013} was essentially ``visual''
(collapse of rescaled plots in log-log),
and the theoretical arguments were reduced to a heuristic derivation, 
the present paper provides a substantial improvement
in support of the validity of a finite-size scaling law
in word-frequency distributions.

\section{Validity of the finite-size scaling law for word-frequency distributions}
{Let us explain Yan and Minnhagen's points \cite{Yan_Minnhagen} in more detail.
They base their analysis on 
%In their Figs. 1 (a) and 1 (b) they compare, for the complete text 
%(with the total length $L=L_{tot}$), 
the empirical
value of the number of types with frequency equal or greater than $k$, 
defined as
$$
N_L(\ge k)= V_L S_L(k) = V_L \sum_{k'=k}^\infty D_L(k'),
$$
%with the vocabulary size $V_L$ for variable $L$, which verifies, by definition,
%$V_L = N_L(\ge 1/L)$. 
where $S_L(k)$ is the empirical complementary cumulative distribution of the frequency
and $N_L(\ge k)$ turns out to be nothing else than the empirical rank
associated to frequency $k$.
In terms of $S_L(k)$ the scaling law 
(\ref{eq:scalingfontclos}) transforms into \cite{Font-Clos2013}
$$
S_L(k) = \frac 1 {V_L} G(k/L), \mbox{ for } k > 10,
$$
under a continuous approximation,
with $G(z)$ a new scaling function directly related to $g(z)$ \cite{Font-Clos2013}.
So, for $N_L(\ge k)$ one has that the scaling law (\ref{eq:scalingfontclos}) can be written as
\begin{equation}
N_L(\ge k) = V_L S_L(k) = G(k/L), \mbox{ for } k > 10.
\label{lachunga}
\end{equation}
Then, in their Figs. 1 (a) and 1 (b) Yan and Minnhagen \cite{Yan_Minnhagen} 
compare this cumulative number evaluated for the complete text,
$N_{L_{tot}}(\ge k)$, 
%(with the total length $L=L_{tot}$),
with the vocabulary size $V_L$ for variable $L$, which verifies, by definition,
$V_L = N_L(\ge 1)$
(note that $L_{tot}$ is the length of the complete text). 
The disagreement between $N_{L_{tot}}(\ge k)$ and $N_L(\ge 1)$
for the same values of the ratio between frequency and length ($k/L_{tot}=1/L$)
makes it clear that the scaling law, for any $L\ne L_{tot}$
and under the form given by Eq. (\ref{lachunga}), 
does not work for the corresponding $k=1$ 
(the \emph{hapax legomena}, 
these are types that appear just once in a text sample of length $L$).
Nevertheless this disagreement does not invalidate the scaling law (\ref{eq:scalingfontclos}) 
for $k>1$.
%
%Note that
%in terms of the cumulative distribution $S_L(k)$ our scaling law 
%(\ref{eq:scalingfontclos}) transforms into \cite{Font-Clos2013}
%$$
%S_L(k) = \frac 1 {V_L} G(k/L), \mbox{ for } k > 10,
%$$
%under a continous approximation,
%with $G(z)$ a new scaling function directly related to $g(z)$.
%So, for $N_L(\ge k)$ we have that 
%$N_L(\ge k) = V_L S_L(k) = G(k/L)$.

Subsequently, 
in their Fig. 2, the same authors \cite{Yan_Minnhagen} plot $N_L(\ge k)$ versus $k/L$
for all $k$ and different $L$ and find indeed deviations with respect the scaling law,
but let us note that these deviations are restricted to $k\le 10$.
Obviously, the reason of these deviations is just that 
the scaling law is expected to hold 
only asymptotically, 
which in practice means $k> 10$ or so 
(see our Appendices I and II, anyhow). 
}

\begin{figure*}
\begin{center}
(a)\includegraphics[width=12cm]{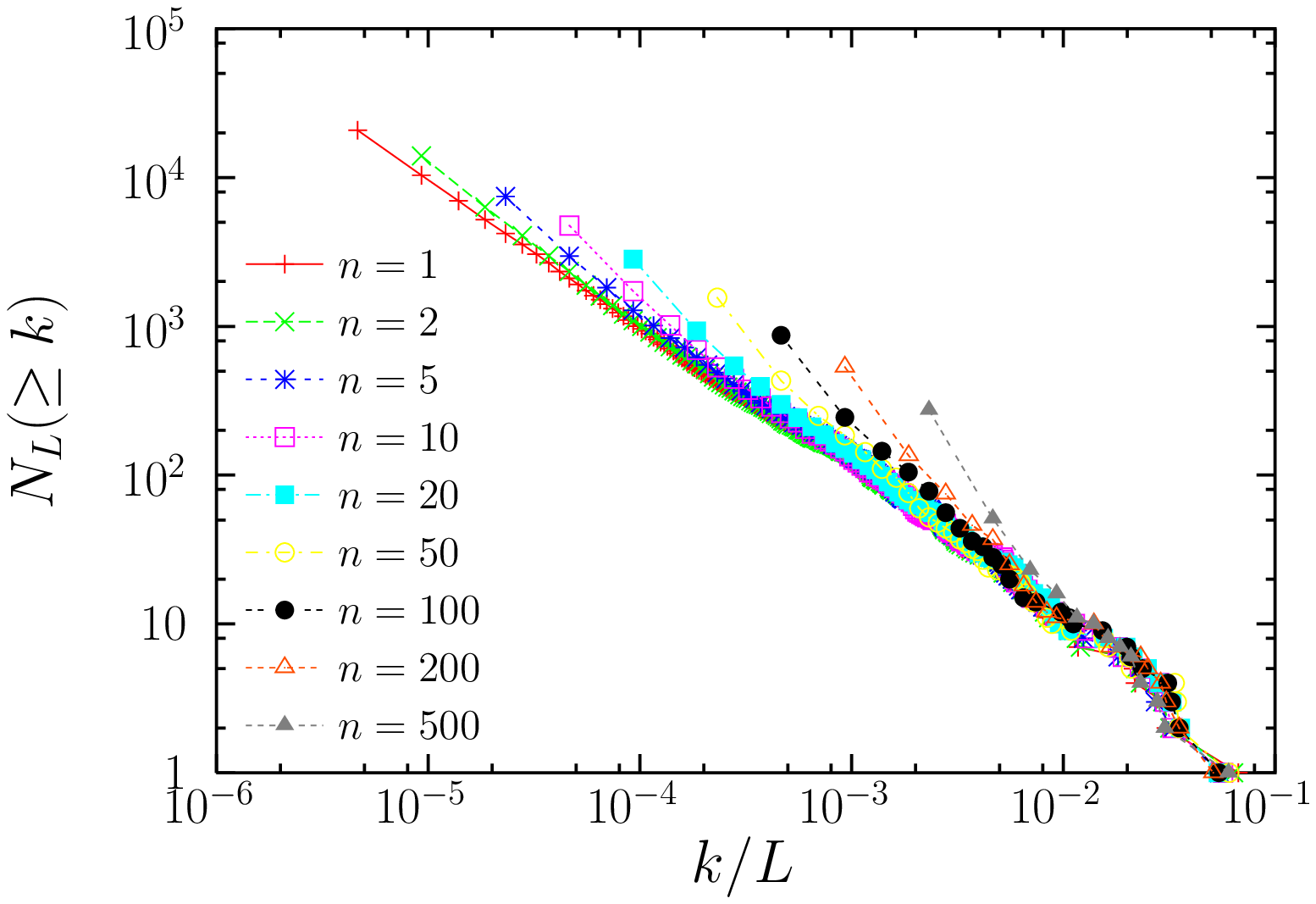}
(b)\includegraphics[width=12cm]{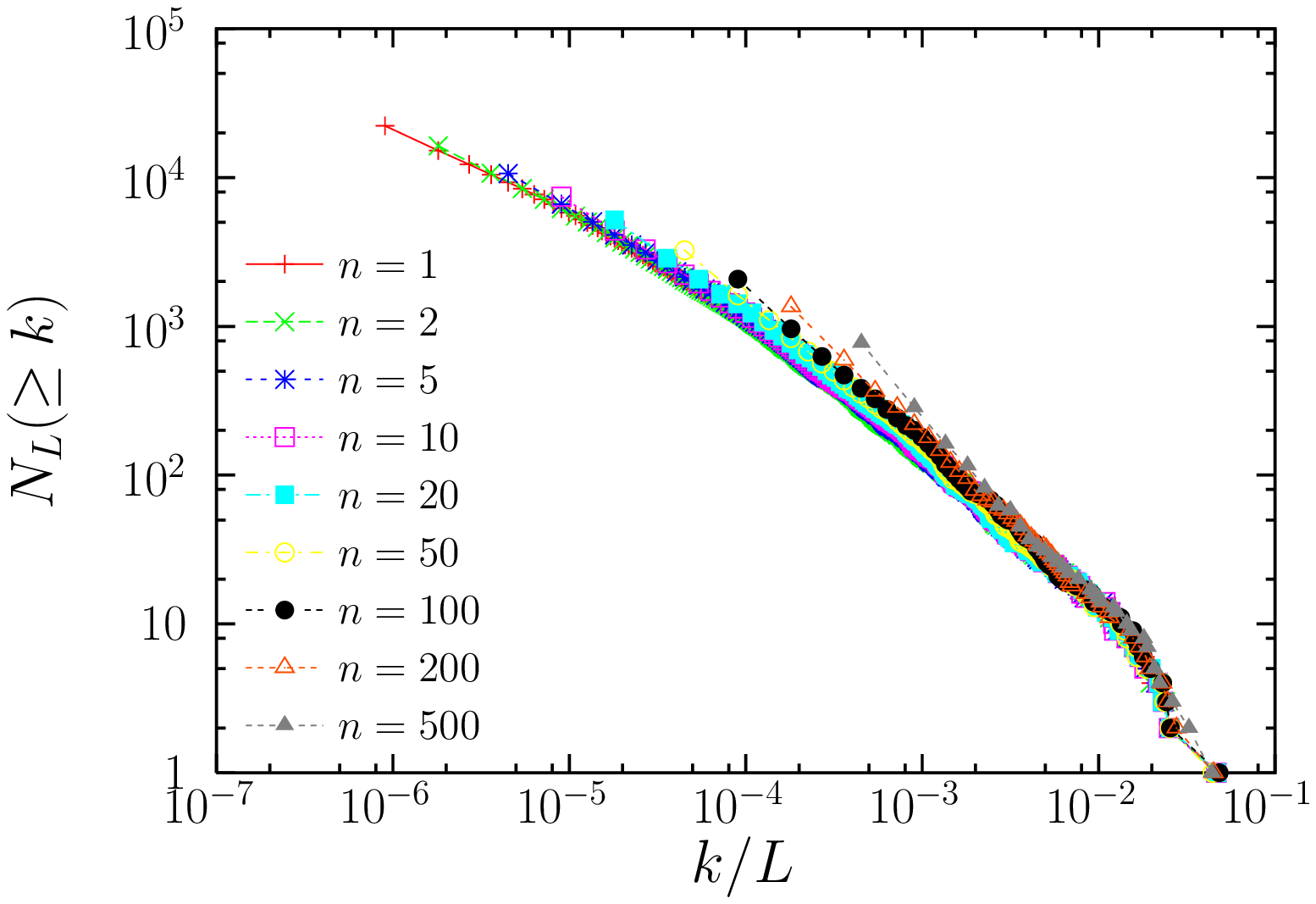}
\end{center}
\caption{
\label{fig_5points}
%{\bf OJO!! La notacion de los captions no es coherente, cambiar en el eje vertical k/L por k!!!
%Y $>$ por $\ge$ !!!!\\}
The total number of words $N_L$ with a relative frequency greater than or equal to $k/L$, for varying $L=L_{{tot}}/n$, with $L_{tot}$ the length of the complete text. 
We have taken the same books as in Ref. \cite{Yan_Minnhagen}, \emph{Moby-Dick} (a) and \emph{Harry Potter} (b), exactly reproducing panels (a) and (b) of Fig.~2 in Ref. \cite{Yan_Minnhagen}, but also including some additional values of $n$. 
%Lines are drawn for all $k$, but symbols are drawn only for $k=1\dots 5$, showing that 
Deviations from the scaling law are always in the regime of very low frequencies, as expected due to discreteness effects
(which are due to the fact that word tokens are discrete). 
}
\end{figure*}

To make our thesis totally clear, in Fig. \ref{fig_5points} 
we present $N_L(\ge k)$ for the same data as in Fig. 2 of Ref. \cite{Yan_Minnhagen}
(i.e., \emph{Moby-Dick}, by Herman Melville, and 
\emph{Harry Potter}, books 1 to 7, by J. K. Rowling),
but adding symbols %for $k=1\dots5$ 
(instead of only lines, as in Ref. \cite{Yan_Minnhagen}). 
It is apparent that even in the extreme case of $500$ equal fragments of the full text, 
the scaling law only fails for very small frequencies. 
%%Naturally, 
We have verified that the scaling law holds for many other texts \cite{Font-Clos2013},
even for \emph{Finnegans Wake}, by James Joyce,
which constitutes an extreme case of experimental literary creation,
yielding an unusual somewhat concave relation between $N_L(\ge k)$ and $k$
(in log-log)
\cite{Kwapien}; in any case, the scaling function $g(z)$ does not care about
concavity or convexity. 
%and very high multifractality \cite{}.

\begin{figure*}
\begin{center}
(a)\includegraphics[width=12cm]{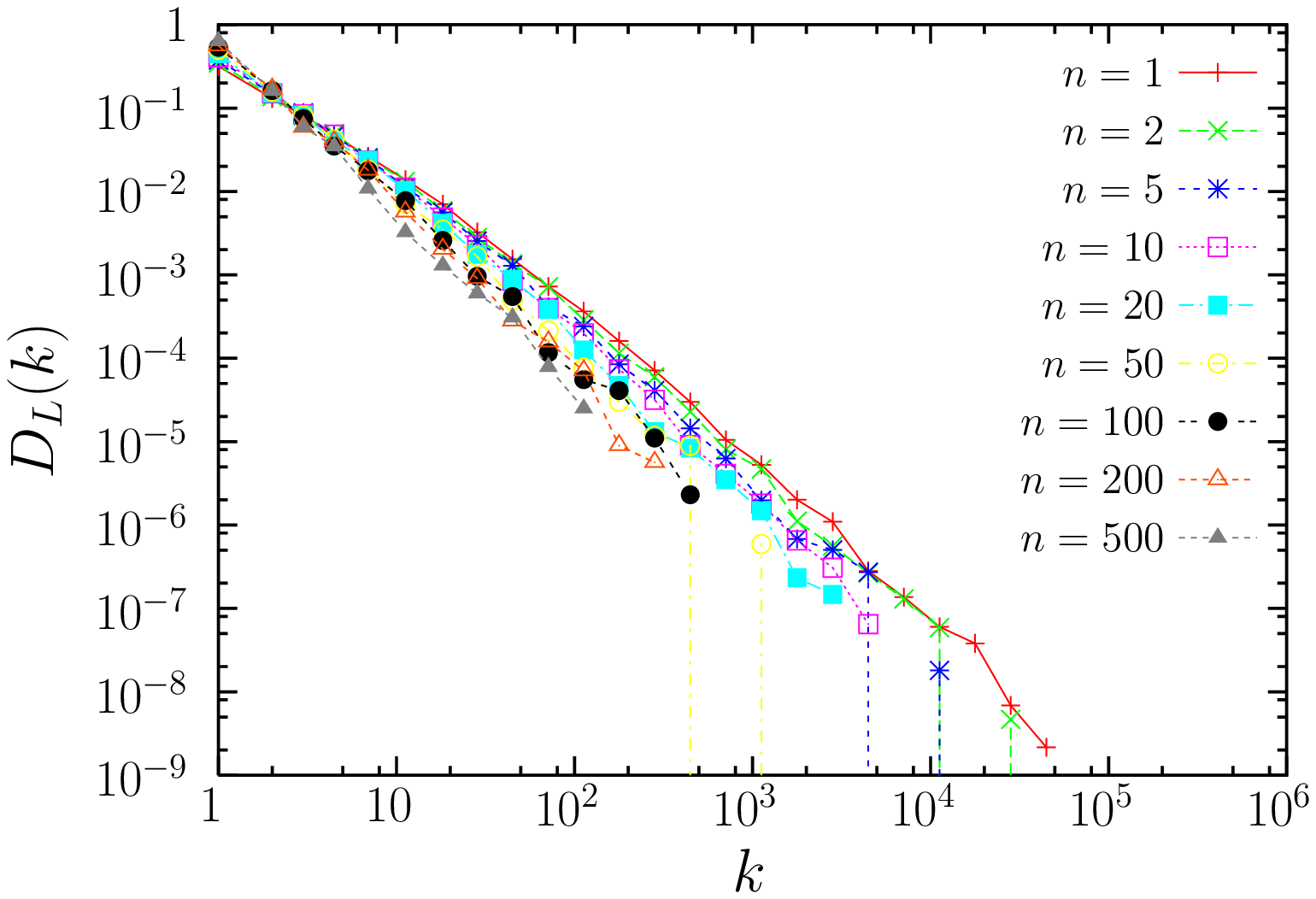}
(b)\includegraphics[width=12cm]{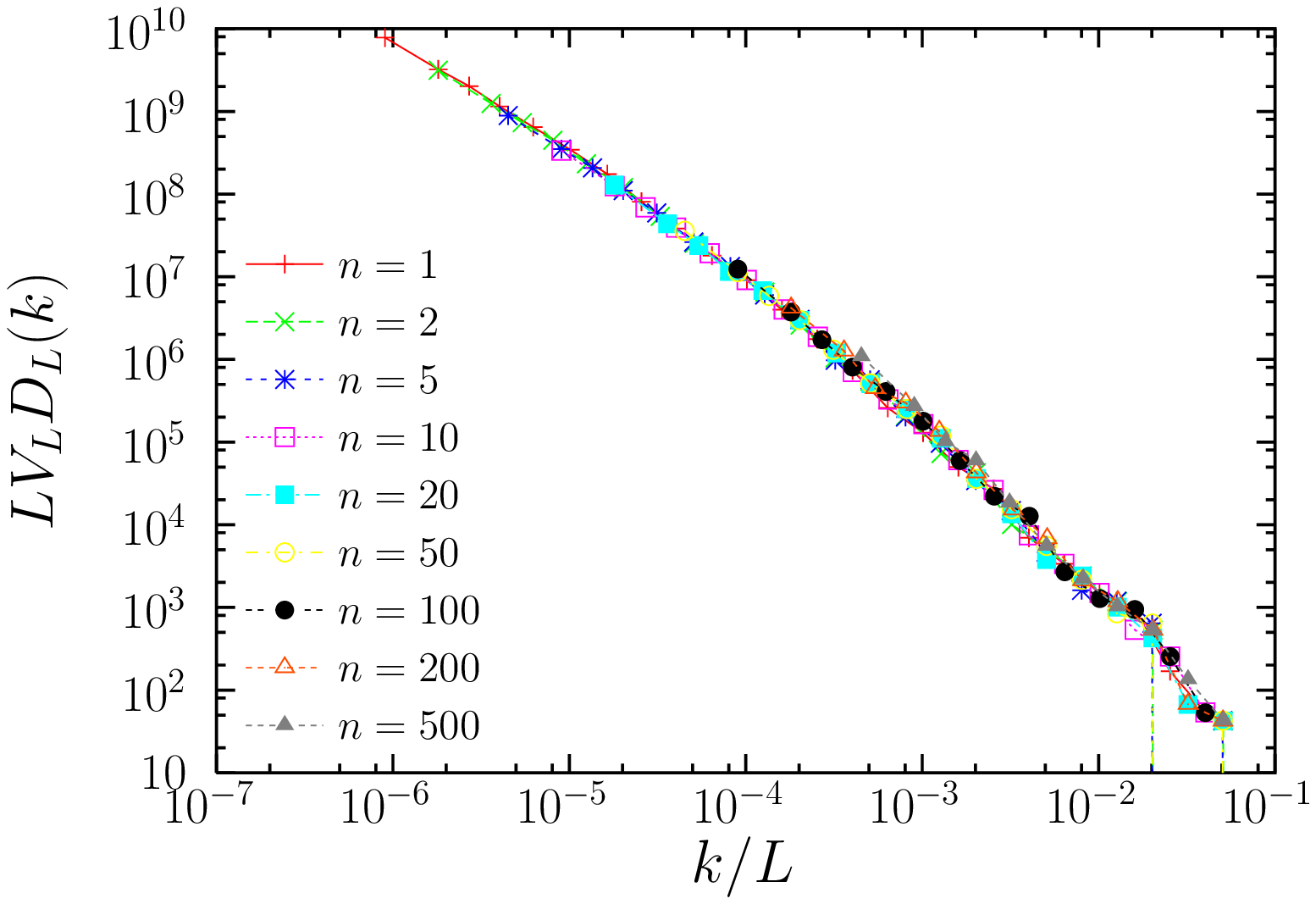}
\end{center}
\caption{
\label{fig_harrypotter}
{(a)} The probability mass function $D_L(k)$ of the absolute frequency $k$, for varying subsets of length $L=L_{{tot}}/n$ of \emph{Harry Potter}, displaying a seeming change of shape. 
{(b)} Same curves, but plotting $D_L(k) L V_L$ versus $k/L$, as proposed in Ref. \cite{Font-Clos2013} and stated here in Eq.~\eqref{eq:scalingfontclos}. All curves collapse into a single, length-independent scaling function $g(k/L)$, in agreement with Eq.~\eqref{eq:scalingfontclos}.
Note the excellent data collapse: even the deviations for small $k$ and negligible for this text.
This is at odds with Eq.~\eqref{eq:scalingsuecos}: a length-dependent exponent in $D_L(k)$, as proposed by Yan and Minnhagen, is not compatible with the data collapse shown in the figure.
}
\end{figure*}

Additionally, in Fig. \ref{fig_harrypotter} we perform the data collapse associated to the scaling law
in terms of the probability mass function $D_L(k)$ for \emph{Harry Potter},
presented in Ref. \cite{Yan_Minnhagen} as a counter-example to the scaling law. 
As it is shown, the collapse is excellent: after proper rescaling 
[Fig. \ref{fig_harrypotter}(b)],
all curves collapse into a single, length-independent function, even for very small frequencies (smaller than 10).
So we can write, for this text, 
$$
D_L(k) \simeq \frac 1 {L V_L} g(k/L), \mbox{ for all } k.
$$
Notice that this is the original form under which 
the scaling law for word-frequency distributions was
presented \cite{Font-Clos2013},
and not the one in Eq. (\ref{lachunga}).
In other words, deviations from the scaling law play an even minor role in the
representation in terms of $D_L(k)$, 
in comparison to the representation in terms of $N_L(\ge k)$.
Finding a functional form for $g(z)$
[whose particular shape in the case of \emph{Harry Potter}
is unveiled in our Fig. \ref{fig_harrypotter}(b)]
is a delicate issue, and it is not our interest here
[in Ref. \cite{Font-Clos2013}, when considering lemmatized texts,
a double power law was proposed for the sake of illustration]; 
nevertheless, in the next section we show that the empirical data,
for each text and different values of $L$,
is compatible with a unique $g(z)$ 
characterized by a power-law tail, with an exponent close to two.

\section{Proper fitting of the power-law tail}

\begin{figure*}
\begin{center}
\includegraphics[width=12cm]{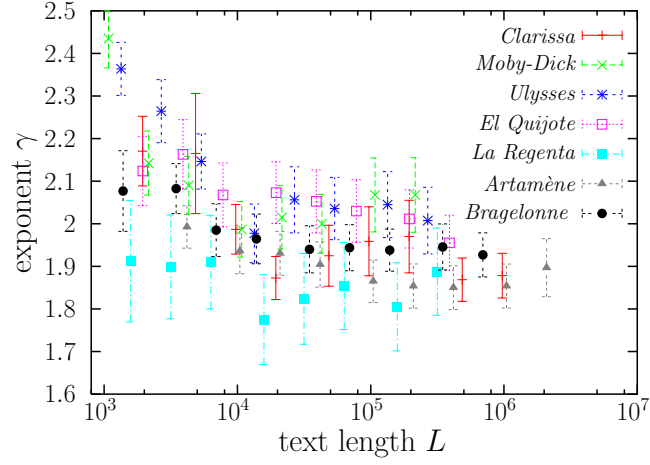}
\end{center}
\caption{
\label{fig_exponents}
Exponents of the power-law tails as a function of text length, 
for seven texts in English, Spanish, and French.
Fits are performer as in Refs. \cite{Corral_Deluca_arxiv,Corral_Boleda}, 
with $k_{cut}$ restricted to the last two decades of the distribution, and
accepted for the $k_{cut}$ that yields $p-$value larger than 0.20
(calculated from 1000 Monte-Carlo simulations).
The error bars correspond to one standard deviation.
Types are defined at the word-lemma-tag level, 
more details on text processing are given in Ref. \cite{Corral_Boleda}.
Note how for $L> 3000$ all exponents are below 2.2, 
and they become quite stable in the range $6000 < L < 2 \cdot 10^6$.
}
\end{figure*}

{Looking at our Fig. \ref{fig_harrypotter}(a),
where $D_L(k)$ is shown with no rescaling, 
one could 
%wrongly 
conclude, as Yan and Minnhagen \cite{Yan_Minnhagen},
that for different text lengths one gets different shapes for $D_k(L)$.
Indeed, a visual inspection of the plot seems to show
different slopes for different $L-$values,
corresponding to different exponents $\gamma(L)$.
However, the data collapse after rescaling in Fig. \ref{fig_harrypotter}(b) 
demonstrates that all distributions have the same shape, 
given by $g(z)$, but at different scales, given by $L$ 
(remember that in log-log a rescaling is seen just as a shift).
And the deviations for $k \le 10$ do not play any relevant role.
%(we quantify this below).
The apparent different slopes of the different curves 
in Fig. \ref{fig_harrypotter}(a) are
a visual artefact caused by the convex (but close to linear) log-log shape
of the curves.
In other words, the larger $L$, 
the more part of $g(z)$ one sees beyond the tail.
As the body of the distribution does not decay as fast as the tail, 
the more part of the body one sees the smaller the apparent
exponent, which is a sort of average between the body and the tail.
This illustrates how a simple replotting under a rescaled form
can unveil a common pattern in distributions given at different scales.
And let us repeat that we are not interested here in providing a parametric model
for this convex shape; for that, just see Ref. \cite{Moreno_Sanchez}.

In order to support their point, 
Yan and Minnhagen \cite{Yan_Minnhagen} fit power laws
[Eq. (\ref{eq:scalingsuecos}) with $c_0=\infty$]
to the word frequency distributions for different $L$,
finding a drift from $\gamma\simeq 3$ for the smallest
fragments of text to $\gamma\simeq 1.6$ for the largest length
(Zipf's law would correspond to $\gamma=2$,
strictly speaking).
The authors do not mention 
%neither 
which fitting method they use, 
nor the fitting range, 
but we can demonstrate that power-law fits 
for the whole range of $k$
in \emph{Moby-Dick} and \emph{Harry Potter} 
are in general rejected after rigorous goodness-of-fit tests, 
no matter if one fits continuous \cite{Corral_Deluca} 
or discrete power laws \cite{Moreno_Sanchez}.
Taking \emph{Harry Potter} and
applying the maximum-likelihood fitting plus the Kolmogorov-Smirnov test
detailed in Ref. \cite{Moreno_Sanchez}, we only get one case
of a non-rejectable
discrete power law defined in the range $k\ge 1$, 
which corresponds to $L=L_{tot}/500$, with an exponent
$\gamma=2.11 \pm 0.04$ and a $p-$value $=0.49$.
For all other lengths of fragments, power laws defined for $k\ge 1$ are rejected
at the 0.05 significance level.

On the contrary, if we fit a power-law tail, which is a power law 
\begin{equation}
D_L(k) \propto \frac 1 {k^{\gamma_2}}
\label{powerlawtail}
\end{equation}
defined only for $k \ge k_{cut}$, 
with $k_{cut}$ a cut-off value of $k$ verifying $k_{cut} > 1$,
we find non-rejectable power laws for all $L$ with stable exponents
when $k_{cut}$ is large enough.
For the case of \emph{Harry Potter} we find that for
$k_{cut} \simeq 0.0009 L$, the exponents $\gamma_2$ turn out to be 
stabilized 
with values very close to two.
%between 2.05 and 2.18. 
So, $g(z)$ has a power law tail, valid for $z=k/L > 0.0009$ and 
with a stable exponent,
at odds with Ref. \cite{Minnhagen2009}'s claims.
Details are available in Table \ref{table_data}.
Figure \ref{fig_exponents} provides more examples of this behavior, 
for different texts.
Although these results are in agreement with Zipf's law
[and in disagreement with the exponential tail represented by Eq. (\ref{eq:scalingsuecos})], 
we are not interested here in the parametric form of the distribution
and only report the stability of the exponents with $L$ as a signature of the
existence of a 
well-defined, $L-$independent
scaling function $g(z)$.
}

\begin{table}[ht]
\caption
{
Results of goodness-of-fit test \cite{Corral_Deluca_arxiv} 
of a discrete power-law distribution $D_L(k) \propto 1/k^{\gamma_2}$
in the range $k \ge k_{cut}$ for \emph{Harry Potter}
(for which $L_{tot}=1108955$). 
Notice the stability of the $\gamma_2$ exponent for 
$k_{cut} \propto L$.
Note also that for $L=L_{tot}/500$ the difference between Yan and Minnhagen's result
%for $k_{cut}=1$ 
(which is $\gamma\simeq 2.4)$ and our result here with $k_{cut}=2$
(which is 2.16) is rather large.
It has been pointed out that inappropriate fitting methods can lead to biased results
\cite{Clauset,Corral_Deluca}.}
\begin{tabular}{| l| c |c| c| c| }
\hline
$L$ & $\gamma_2$ & $k_{cut}$ & 
%$N_L(\ge k_{cut})$& 
$V_L$ &$p-$value\\
\hline
$L_{tot} $ & 2.06 $\pm$ 0.09 & 1024 & %149 & 
22276 &0.677 \\
$L_{tot}/2$ & 2.05 $\pm$ 0.09 & 513 & %148 & 
16361 &0.504 \\ 
$L_{tot}/5$ &2.09 $\pm$ 0.09 & 205 & %153 & 
10658 &0.862 \\
$L_{tot}/10$ & 2.09 $\pm$ 0.09 & 103 & %155 &
7431 & 0.263 \\
$L_{tot}/20$ & 2.18 $\pm$ 0.09 & 52 &%169 &
5186 & 0.819 \\
$L_{tot}/50$ & 2.14 $\pm$ 0.08 & 20 & %183 & 
3240 & 0.554 \\
$L_{tot}/100$ & 2.17 $\pm$ 0.09 & 11 & %180 & 
2079 & 0.708 \\
$L_{tot}/200$ & 2.14 $\pm$ 0.08 & 5 & %219 & 
1353 & 0.903 \\ 
$L_{tot}/500$ & 2.16 $\pm$ 0.07 & 2 & %285 & 
774 & 0.683 \\ 
\hline
\end{tabular}
\label{table_data} 
\end{table}

%Por que crece sistematicamente el numero de types en el rango power law
%si k/L esta fijado??? Es por lo mismo que la ley de escala no se cumple
%para k's pequenyos? En este caso la desviacion es grande!
%De 149 a 285 

\section{Testing of the scaling hypothesis}

A more direct and non-parametric way to test the existence of scaling is to use the two-sample Kolmogorov-Smirnov test \cite{Press}, which compares two data sets
under the null hypothesis that both of them come from the same population,
and therefore have the same underlying theoretical distribution (which is unknown
and remains unknown after the test).
But in the case of scaling we are not dealing with the same distribution,
but with distributions which have the same shape at different scales,
i.e., distributions that are the same except for a scale parameter; 
then, rescaling the distributions by their scale parameter
would lead to the same distributions (under the null hypothesis that
scaling holds).
This procedure to test the fulfillment of scaling
has been used before for continuous distributions 
\cite{Corral_test,Lippiello_Corral}.

The Kolmogorov-Smirnov test is probably the best accepted test for comparing
continuous distributions, but word-frequency data are discrete,
and after rescaling become discretized over different sets
(as the scaling factors $L$ of the two distributions can be very different, in general).
So, our first step, in order to avoid this problem is to approximate the
discrete empirical distributions by continuous ones, 
by adding to each frequency a random term, in this way
$k \rightarrow k + u$ (where $u$ is a uniform random number between
$-0.5$ and $0.5$).
Although there are more sophisticated ways to continuize the distributions, 
this one uses no information from the data 
(except that the $k$'s are natural numbers).
The second step is to remove small frequencies 
(remember that scaling is a large-scale property)
in our case we remove values of $k$ below 4.
Then, the third step is to perform the rescaling
$$
k \rightarrow \frac{\langle k \rangle k}{\langle k^2 \rangle},
$$
where the moments of $k$ are the original empirical ones
(calculated for the discrete distribution).
In a simple case (with no power laws involved \cite{Corral_test,Lippiello_Corral})
we would have rescaled just by the mean $\langle k \rangle$;
in this case the rescaling is a bit more involving \cite{Peters_Deluca,Corral_csf}.
Notice that this rescaling is totally equivalent to divide $k$ by $L$,
as shown in another section below;
nevertheless, our choice is more general and makes the rescaling
applicable when the data does not come from a text.

\begin{figure*}
\begin{center}
\includegraphics[width=12cm]{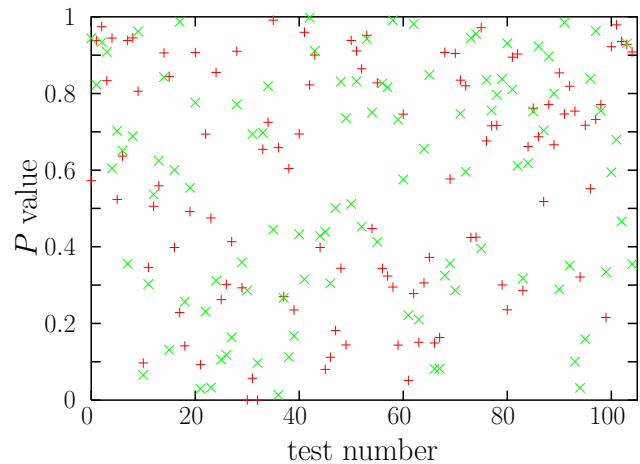}
\end{center}
\caption{
\label{KStests}
$P-$values of two-sample Kolmogorov-Smirnov tests
for word-frequency distributions rescaled as explained in the text.
One of the samples has text length $L$ and the other $L'$,
ranging from $L_{tot}/50$ to $L_{tot}$.
One of the data sets shown corresponds to testing 
the first fragment of length $L$ with the same fragment of length $L'$,
for the other data set other fragments are chosen.
The texts are the same seven used in the previous figure.
$P-$values below 0.05 should lead to the rejection of the 
scaling hypothesis with a 0.95 confidence level.
}
\end{figure*}

Once these three steps have been done, 
the two-sample Kolmogorov-Smirnov test \cite{Press} 
is performed for all pairs of samples
given by different $L$, 
restricting the samples to a common support, 
i.e., a common minimum value is taken as the minimum value 
of the sample with the smallest $L$
(which has the largest minimum when the frequencies are rescaled).
Figure \ref{KStests} shows the $P-$value of this test for several texts and different divisions
of the texts, up to $L_{tot}/50$.
The fact that the $P-$value appears as uniformly distributed between zero and one
is an indication that the scaling null hypothesis holds.

%ESTOS FACTORES LOS USABAMOS TAMBIEN
%EN LOS PAPERS VIEJOS 2009, 2012??
%NO!! Reescalabamos por la media!!!!, no???
%Si, era mas facil segun recuerdo, 
%ya esta explicado

\section{Relative errors of the scaling law}

Although the proper way to compare statistical distributions
is by means of statistical tests (as done in the previous section),
{Yan and Minnhagen \cite{Yan_Minnhagen} use instead relative errors.
They show numbers for 
the relative error provided by the scaling law, 
and compare it with the error of the so-called random-group-formation
hypothesis. 
%We explain why their comparison is very unfair.
We explain why their comparison is not 
appropriate
%fair.
First, for the scaling law, the empirical values of $N_L(\ge k)$ are compared
for fixed ratio $k/L$ with $N_{L_{tot}}(\ge k)$
and the errors are claimed to be large. 
%The errors are claimed to be large, but for instance, 
%dividing the text in 20 parts ($L=L_{tot}/20$) and considering $k> 5$
%we can see in 
%Yan and Minnhagen's Fig. A.3 (a) and (b) that
%the relative error of the scaling law is alwais below 0.3.
Second, for the random group formation the error is claimed much smaller, 
but in this case the empirical data $N_L(\ge k)$ are compared 
with random samples of the same length $L$,
and not with a distribution of a different length.
It is obvious that this procedure has to yield better results, 
and this constitutes a totally biased comparison.

But further, the errors provided by Yan and Minnhagen \cite{Yan_Minnhagen} for the scaling law are
inflated. 
Our Fig. 5 shows the relative difference or error between the true value $N_L(\ge k)$ and the value
approximated by the scaling law, $N_{L_{tot}}(\ge k')$, with $k'=L_{tot} k/L$, which is
$$
\varepsilon_L(k) =
\frac {N_{L_{tot}}(\ge k') - N_{L}(\ge k)}
{N_{L}(\ge k)}.
$$
Note that, in general, the replacement in the denominator of ${N_{L}(\ge k)}$
by ${N_{L_{tot}}(\ge k')}$ inflates the reported error, as, when there are deviations, 
this number is systematically below ${N_{L}(\ge k)}$.
The results of our analysis (Fig. 5) show 
that the errors are not as big as reported by Yan and Minnhagen \cite{Yan_Minnhagen}.
Dividing \emph{Harry Potter} in up to 20 parts,
the relative error provided by the scaling law is almost always below 0.2, 
with the remarkable exception of the case
$k=1$ for $L=L_{tot}/20$.
Dividing the text into smaller parts yields that the relative error
is always below 0.3 for $k>10$.
But the error for small $k$ is further reduced
if one uses for comparison the probability mass function $D_L(k)$
instead of $N_L(\ge k)$.
Remember that the original form of the scaling law was reported for $D_L(k)$
and not for $N_L(\ge k)$.
Our Fig.~\ref{fig_harrypotter}(b) speaks for itself.

\section{Scaling of moments from the generalized central-limit theorem, 
Heaps' law, and relation with the scaling law}

We start this section dealing with
a distribution $D_L(k)$ 
%of type frequency 
that has a power-law tail
%regime, 
%as in Eq. (\ref{powerlawtail}), 
with an exponent in the range $1 < \gamma_1 < 2$.
%We could have considered that the exponent is $\gamma$, 
%as in Eq. (\ref{powerlawtail}), but we pretend to be more general.
We consider the moments $\langle k \rangle$ and $\langle k^2 \rangle$
not as the moments of the theoretical distribution (which would be equal to infinity)
but as the moments of a finite sample, 
whose size is just the size of the vocabulary $V_L$ (by definition); that is,
$$
\langle k \rangle = \frac 1 {V_L} \sum_{i=1}^{V_L} k_i
\mbox{ and }
\langle k^2 \rangle = \frac 1 {V_L} \sum_{i=1}^{V_L} k_i^2.
$$ 
%(note that these empirical moments will depend on sample size $V_L$
%and therefore on $L$).

Due to the power-law behavior for large $k$,
the generalized central limit theorem \cite{Bouchaud_Georges,Corral_csf} 
allows one to obtain the scaling
properties of these sums, assuming that the individual frequencies
are independent (or weakly dependent).
Indeed, 
%due to the power-law distribution, 
$\sum_{i=1}^{V_L} k_i$ does not scale linearly with $V_L$ but
superlinearly, as
$\sum_{i=1}^{V_L} k_i \propto V_L^{1/(\gamma_1-1)}$.
Moreover, if $k$ has a power-law tailed distribution with exponent $\gamma_1$
so does $k^2$, but with exponent $\gamma_1'$ fulfilling 
$\gamma_1'-1 =(\gamma_1-1)/2$ 
(and in the range $1<\gamma_1'<2$)
and then
the generalized central-limit theorem also applies to $k^2$, to give
$\sum_{i=1}^{V_L} k_i^2 \propto V_L^{2/(\gamma_1-1)}$.

On the other hand, we can also use the exact result
$\sum_{i=1}^{V_L} k_i =L$ (the definition of text length), 
from which we obtain the classical Heaps' law
(called also Herdan's law in llinguistics)
\cite{Baeza_Yates00,Kornai2002,Lu_2010,Serrano,Font-Clos2013},
\begin{equation}
V_L \propto L^{\gamma_1-1},
\label{Heaps}
\end{equation}
and therefore the moments fulfill
\begin{equation}
\langle k \rangle = L^{2-\gamma_1}
\mbox{ and }
\langle k^2 \rangle = L^{3-\gamma_1} 
\label{scalingofmoments}
\end{equation}
(and, in general, $\langle k^m \rangle = L^{m+1-\gamma_1}$).

This result is 
%what one obtains when the distribution of $k$ fulfills 
%COMPATIBLE WITH!!!
compatible with
a scaling law of the form 
\begin{equation}
D_L(k)= \frac 1 {L^{\gamma_1}} g\left(\frac k L\right).
\label{otraleydeescalamas}
\end{equation} 
%where notice that $L$ is the scale parameter of the distribution.
The case considered in the literature \cite{Christensen_Moloney,Corral_csf}
assumes that $g(z)$ has an intermediate power-law decay with exponent $\gamma_1$ 
followed by a much faster decay (exponential or so) for the largest $k$'s.
The pure power-law tail considered above is 
included in this framework when $g(z)$ goes to zero abruptly, 
transforming the pure power law into a truncated power law.
Indeed, if the power law is truncated at $k_{max}$,
using a continuous approximation 
%and substituting the scaling law for $D_L(k)$ into the expression for 
%$\langle k^m \rangle$ 
we get
$$
{\langle k^m \rangle} 
=\int_1^\infty k^m D_L(k) dk 
\simeq \int_1^{k_{max}} k^m D_L(k) dk 
$$
\begin{equation}
= 
L^{m+1-\gamma_1} \int_{1/L}^{k_{max}/L} z^m g(z) dz
\sim L^{m+1-\gamma_1},
\label{derivation}
\end{equation}
because (for $m> \gamma_1-1$) the integral tends to a constant when $L$ is large, 
taking into account that $k_{max}$ is 
the maximum of $k$ from a sample of size $V_L$
and scales in the same way as $\sum k_i$, 
i.e., as $L$.
To see this one can just calculate, for any $P$, the percentiles $k_P$ of the distribution
of the maximum of $V_L$ frequencies, 
which verify $[1-S_L(k_P)]^{V_L}=P$.
Substituting a power law for $S_L(k)$ (with exponent $\gamma_1-1$)
we get
$$
k_P \propto \frac 1 {(1-P^{1/V_L})^{1/(\gamma_1-1)}} \simeq
\left(\frac {V_L}{-\ln P}\right)^{1/(\gamma_1-1)} \propto L.
$$
So, as all the percentiles of the maximum scale with $L$,
the distribution of the maximum scales with $L$ too
[we arrive at the last result using that
$P^{1/V_L} = e^{(\ln P)/V_L} \simeq 1 +(\ln P)/V_L$,
valid for large $V_L$, and also Heaps' law (\ref{Heaps})].

Therefore, a power-law tail for the distribution of frequencies, 
with exponent $\gamma_1$, is somehow equivalent to 
an upper truncated power-law tail, 
with an effective cut-off $k_{max} \propto V_L^{\gamma_1-1} \propto L$,
which makes that the features of the distribution at the largest $k$ have to scale with $L$.
The scaling law (\ref{eq:scalingfontclos}) follows directly from 
Eq. (\ref{otraleydeescalamas})
%%$D_L(k) =L^{-\gamma_1} g(k/L)$, 
using Heaps' law (\ref{Heaps}),
although notice that the version of the law given by Eq. (\ref{eq:scalingfontclos})
is non-parametric, in the sense that the value of the exponent $\gamma_1$
does not appear in the law 
(which is good if the determination of the exponent contains errors).

Moreover, as a by-product we obtain another form for the scaling law,
\begin{equation}
D_L(k) = \frac{\langle k \rangle^3}{\langle k^2 \rangle^2} \,
g\left( \frac{k \langle k \rangle}{\langle k^2 \rangle}\right),
\label{elscalingdelosmomentos}
\end{equation}
using the scaling (\ref{scalingofmoments}) of the moments with $L$
and the scaling law,
with the scaling function $g$ being the same as before, except
for proportionality factors.
This scaling has been used previously for self-organized critical phenomena 
but under different conditions \cite{Peters_Deluca}.
Remember that here the moments are not those of the theoretical distribution
but the ones corresponding to a sample of size $V_L$. 
The equivalence of both scaling laws, 
Eqs. (\ref{eq:scalingfontclos}) and (\ref{elscalingdelosmomentos}),
is empirically shown in Fig. \ref{figL2V} by means of the proportionality
between $\langle k^2 \rangle$ and $L^2/V_L$.
%Notice that, from this perspective, the scaling function effectively includes a sharp
%upper cutoff at $k_{max}\propto L \propto V_L^{\gamma_1-1}$;
%in other words, for finite samples, in practice, the power-law tail is truncated at the 
%maximum frequency $k_{max}.$

However, real distribution of frequencies are not well described 
at the largest frequencies 
by scaling functions $g(z)$ that decay either exponentially or abruptly 
(i.e., are sharply truncated),
as shown in Table \ref{table_data} and Fig. \ref{fig_exponents}. 
Instead, we expect that the tail of $g(z)$ 
[and therefore the tail of $D_L(k)$] is another power law,
with an exponent $\gamma_2> \gamma_1$.
Remarkably, this framework is also described by 
the scaling law (\ref{otraleydeescalamas})
and the scaling of moments (\ref{scalingofmoments}),
being the key point that one can change the upper limit of the
integral from infinite to $k_{max}$,
and $k_{max}$ still scales linearly with $L$, 
so the derivation is the same as in Eq. (\ref{derivation}).

In order to support empirically the fulfillment of a scaling law of the 
form given by Eq. (\ref{otraleydeescalamas}) we follow the approach presented
in Ref. \cite{Deluca_npg}.
If such a scaling law holds, the distance between the different rescaled distributions
in log-scale 
$$
(\ln k_i -\delta \ln L, \ln D_L(k_i) + \gamma_1 \ln L)
$$
should be minimum when the right values of the exponents are substituted.
Notice that we have introduced an extra exponent $\delta$, which we expect
becomes equal to 1.
We proceed by minimizing such distances as a function of the exponents 
$\delta$ and $\gamma_1$, 
resulting in values of $\delta$ very close to 1 indeed
and values of $\gamma_1$ in the range 1.5 to 1.95, 
when different texts are used,
see Table \ref{laotratabla}.

\begin{table}[ht]
\caption
{Exponents $\delta$ and $\gamma_1$ obtained after the minimization of the
distance between the rescaled word-frequency distributions, optimizing the data collapse,
for 7 different texts.
Text length $L$ varies from $L_{tot}$ to $L_{tot}/10$.
Different seeds are used in the algorithm in order to avoid local minima.
}
\begin{tabular}{|l|l|c|c|}
\hline
Text and author & Language & $\delta$ & $\gamma_1$ \\
\hline

\emph{Clarissa} by S. Richardson & English & 0.96 & 1.50\\
\emph{Moby-Dick} by H. Melville & English & 1.05 & 1.87\\
\emph{Ulysses} by J. Joyce & English & 1.04 & 1.94\\
\emph{El Quijote} by M. de Cervantes & Spanish & 0.96 & 1.64\\
\emph{La Regenta} by L. A. Clar\'{\i}n & Spanish & 0.86 & 1.52\\
\emph{Artam\`ene} by Scud\'ery siblings& French & 1.04 & 1.63\\
\emph{Le Vicomte de Bragelonne} & French & 0.96 & 1.58\\
by A. Dumas (father)& & & \\

\hline

\hline
\end{tabular}
\label{laotratabla} 
\end{table} 

\section{Discussion and conclusions}

As an important remark,
we want to clarify that we are not against the so-called random group formation hypothesis \cite{Baek2011}, 
as in some previous research we have made use of randomness to explain real texts \cite{Font_Clos_Corral}.
Our conclusion was that real texts are not random, but the first appearance of 
a word is close to random, so the word frequency distribution (related to Zipf's law) 
and the type-token growth curve (related to Heaps' law) remain the same
for real texts and for random versions of them.
The reason is that the word frequency distribution is independent on word order
and the type-token growth curve only depends on the first appearance of a word.
Other properties of real texts are different from those of random texts, 
as inter-appearance distances \cite{Font_Clos_Corral,Altmann_Motter,Corral_words}.

}

Summarizing, the empirical facts are clear: 
a finite-size scaling law gives a very good approximation for the 
distribution of word frequencies
for different fragments of text of length $L$.
% in the range $k > 10$. 
The shape of $D_k(L)$ is the same for all $L$,
and it is only a scaling factor proportional to $L$ what makes the difference
for different $L$.
{It is the parametric proposal of Ref. \cite{Minnhagen2009}
which is not well supported by solid statistical testing.}
In any case, if the theory held by Yan and Minnhagen 
\cite{Yan_Minnhagen}
is valid, then it must contain in some limit the scaling law.
If not, the theory is irrelevant for real texts.
%(perhaps because it is based on wrong postulates). 

In conclusion, we show how the sort of scaling arguments
usual in statistical physics, 
and in particular finite-size scaling (\ref{eq:scalingfontclos}) 
can describe complex processes
much better than parametric formulas (\ref{eq:scalingsuecos}).
Finally, in order to avoid misunderstandings, let us state that
although curve fitting is a very honorable approach in science
(when done correctly \cite{Corral_Deluca_arxiv,Corral_Deluca,Moreno_Sanchez}), 
a scaling approach has nothing to do with that
%contrary to Yan and Minnhagen's claims 
\cite{Yan_Minnhagen}. 

%{OJO!!! si la ley de escala se cumpliera exactamente en D, 
%no se cumpliria en S por efectos discretos, ni vice-versa.
%Cual el la ley buena, i.e., cual es mas fundamental??}

\section{Acknowledgements}

We are grateful to R. Ferrer-i-Cancho for drawing our attention
to Ref. \cite{Minnhagen2009}, 
and to G. Boleda for facilitating the beginning of this research.
Yan and Minnhagen's criticisms have allowed us
to explain in much more detail the validity of our scaling law
and to arrive to the new results presented here.
Research projects in which this
work is included are FIS2012-31324 and FIS2015-71851-P, 
from Spanish MINECO, and 2014SGR-1307, from
AGAUR.

\section*{Appendix I}

We explain here the difference between a power law and a scaling law,
and how scaling laws in statistical physics usually only hold asymptotically.
Let us start with a scale transformation. This is an operation that stretches and/or
contracts a function, i.e.,
$$
T[f(x,y)] = c f(x/a,y/b),
$$
where $f(x,y)$ is a (in this example bivariate) real function,
$a,b,c$ are constant and positive scale factors, and $T$ is the scale transformation.
If we ask the question about which functions are invariant under scale transformations
(i.e., which functions do not change when are stretched and/or compressed) we find that
a solution is
\begin{equation}
f(x,y)=\frac 1 {x^\beta} g\left(\frac y {x^\alpha}\right),
\label{scalinglaw}
\end{equation}
with $\alpha=\ln b / \ln a$ and $\beta=-\ln c / \ln a$,
and $g$ an arbitrary function, 
called scaling function
(we also consider $x>0$).
Moreover, if we look for a solution valid for any real value of $a >0$,
the previous solution (\ref{scalinglaw}) turns out to be the only solution \cite{Christensen_Moloney}.
One refers to Eq. (\ref{scalinglaw}) as a scaling form or a scaling law.

Notice that a power law is a special case of scaling law, 
just taking the arbitrary scaling function $g$ to be a constant $C$.
In fact, in one dimension (i.e., for univariate functions $f(x)$)
the only scaling laws (the only scale-invariant functions for any value of the scale factor $a$)
are the power laws \cite{Newman_05,Corral_Lacidogna}, so $f(x)=C/x^\beta$.
Although the terms ``scaling law'' and ``power law'' are sometimes taken as synonym of each other, it is clear that
they are only equivalent for univariate functions. 
For bivariate (and multivariate) functions one needs to be more careful
in distinguishing both concepts (as we do here).

Let us stress that scaling is a fundamental pillar of 20th-century statistical physics 
\cite{Stanley_rmp}.
In our case, we propose that a (bivariate) scaling law holds for $D_L(k)$, 
so, we identify $k=y$, $L=x$, and $D_L(k)=f(x,y)$
and assume Heaps' law for the usual scaling law to hold
(more details in Ref. \cite{Font-Clos2013}).
Alternatively, we may identify $N_L(\ge k)$ not with $D_L(k)$ but with $f(x,y)$
with no necessity of using Heaps' law.
In any case this not necessarily implies that the scaling function
$g$ has a power-law shape.
We do not care here about the functional form of $g$,
this is just the shape shown in Fig. \ref{fig_harrypotter}(b)
(for the particular book under consideration there).

We provide in Fig. \ref{figappendix} a practical example of 
how scaling laws in statistical physics hold usually only for large $x$ and $y$
(i.e., large $L$ and $k$).
We display the rescaled size distribution $D_L(k)$ of a critical Galton-Watson branching process \cite{Corral_FontClos}
with its number of generations bounded by a finite $L$
and with offspring distribution given 
by a binomial distribution with 2 trials.
This process is totally equivalent to percolation in the Bethe lattice
\cite{GarciaMillan}.
The figure shows the deviation from the scaling law for $k\le 10$,
but nevertheless it has been proved analytically that
finite-size scaling holds in this system \cite{GarciaMillan}.
Ironically, in this case the scaling function
is well approximated by the function proposed by 
Bernhardsson et al. \cite{Minnhagen2009}
[our Eq. (\ref{eq:scalingsuecos})],
but with a constant exponent $\gamma_{}=\beta/\alpha=3/2$.

\section*{Appendix II}

Let us see how discreteness effects alter scaling.
Naturally, the discrete nature of word-frequency distributions
comes from the fact that the fundamental unit is the count
of word tokens.
If we assume that scaling holds for all $k$,
even for $k=1$ and $k=2$,
the finite-size scaling law, under the form given 
by Eq. (\ref{lachunga}), implies that
$$
N_L(\ge 2) = N_{L/2}(\ge 1),
$$
and we can relate this to the size of vocabulary for each text length,
so
$$
V_L - n_L^s(1)=V_{L/2},
$$
where $n_L(k)$ counts the number of types with frequency 
(exactly equal to) $k$, 
and the superscript $s$ denotes that we are under the scaling hypothesis.

On the other hand, for a random text,
$V_{L/2}$ can be calculated 
from $V_L$ as
$$
V_{L/2}= V_L - n_{L/2}(0)
= V_L - \sum_{k\ge 1} h_{0,k} n_L(k),
$$
where $h_{0,k}$ gives the probability of getting $0$ tokens
of a certain type when a fragment of text of length $L/2$
is taken from a text of length $L$ in which the same type has
frequency $k$, see Eq. (5) of Ref. \cite{Font-Clos2013}.
Comparing both equations for $V_{L/2}$ we get 
$n_L^s(1)=\sum_{k\ge 1} h_{0,k} n_L(k)$.
But using that $h_{0,k} < 1/2^k$ (see below) and $n_L(k)< n_L(1)$ for $k>1$
(from empirical evidence, see Fig. 2(a) for instance)
we arrive at
$$
n_L^s(1)=
n_{L/2}(0)=
\sum_{k\ge 1} h_{0,k} n_L(k)
< n_L(1)
\sum_{k\ge 1} \frac 1 {2^k} = n_L(1)
$$
(extending the sum to infinite).
Thus, the scaling hypothesis yields, for a random text and for $k=1$,
less types than it should.
This can be seen looking carefully at some of the plots in 
Fig. 2 of Ref. \cite{Font-Clos2013}, but not in Fig. 3 there or in Fig. 2(b) here, 
as the deviations are rather small
[just notice that the empirical $D_L(k)$
is proportional to $n_L(k)$].

The fact that $h_{0,k} < 1/2^k$ comes from the fact
that $h_{0,k}$ is given by the hypergeometric distribution
(as we assume we take tokens from the larger text with no replacement), 
and then,
\begin{alignat}{1}
h_{0,k} &= 
{\binom{k}{0}\binom{L-k}{L/2-0}}\Bigl/\Bigr.{\binom{L}{L/2}}\\
&=
\left[
\frac{(L-k)!}{(L-k-L/2)!} \right]
\Bigl/\Bigr.
\left[
\frac{L!}{(L/2)!}
\right]\\
&=
\left[
\frac {(L/2)!}{(L/2-k)!}\right]
\Bigl/\Bigr.
\left[
\frac
{L!}{(L-k)!}
\right]\\&=
\frac{
(L/2)(L/2-1)\dots(L/2-k+1)
}{
L(L-1)\dots(L-k+1),
}\\
&=
\prod_{j=0}^{k-1} 
\left(
\frac{L/2-j}{L-j}
\right),
\end{alignat}
where all factors are smaller than $1/2$,
except the one for $j=0$.
This yields $h_{0,k} < 1/2^k$.
In this way we show how discrete effects break scaling
for the lowest frequencies, 
but, as can be seen in the plots, this effect is very small.

%%\addcontentsline{toc}{chapter}{Bibliography}
%%\vspace*{-3mm}
%%%\bibliography{D:/Dropbox/projects/words_ramon/p1_lemmas/biblio.bib}
%\bibliography{../../../../../projects/words_ramon/p1_lemmas/biblio,biblio}
%%%\bibliography{biblio}
%\bibliographystyle{unsrt}

\begin{figure*}
\begin{center}
\includegraphics[width=12cm]{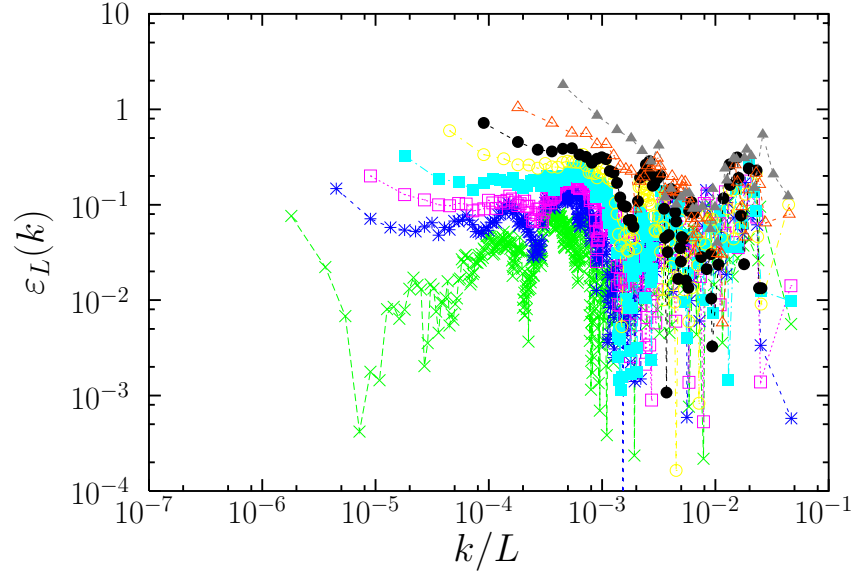}
\end{center}
\caption{
%{\bf OJO!!! Los coloringos aqui son un poco raro, 
%quedarian mejor si se correspondieran con las otras figuras, 
%pero tampoco es demasiado importante...\\}
Relative error 
$[N_{L_{tot}}(\ge k') - N_{L}(\ge k)]/{N_{L}(\ge k)}$ 
of the approximation to $N_{L}(\ge k)$ given
by the scaling of the distribution of the whole text $N_{L_{tot}}(\ge k')$,
with $k'=k L_{tot} /L$, for \emph{Harry Potter}. 
Symbols correspond to the same values of $n$ as in the previous figures.
\label{figerror}
}
\end{figure*}

\begin{figure*}
\begin{center}
\includegraphics[width=12cm]{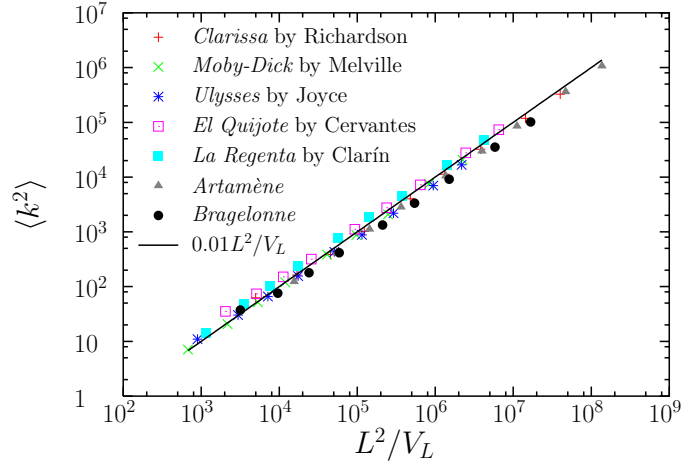}
\end{center}
\caption{
\label{figL2V}
Linear relation between the second moment of the distribution of word frequencies
and the ratio of the squared text length with vocabulary size,
for diverse texts used in the previous figures \cite{Corral_Boleda}.
A line with a linear coefficient 0.01 is shown for comparison.
}
\end{figure*}

\begin{figure*}
\begin{center}
\includegraphics[width=12cm]{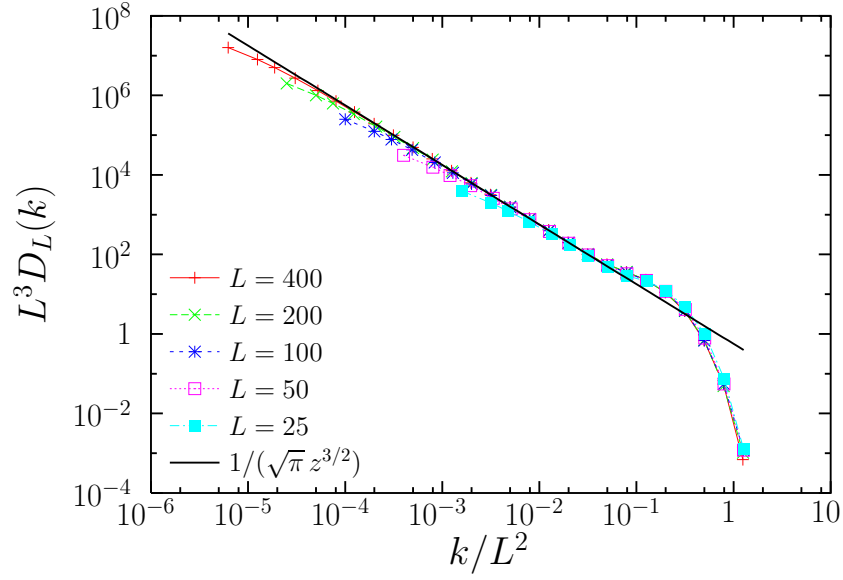}
\end{center}
\caption{
\label{figappendix}
Rescaled size distribution $D_L(k)$ as a function of rescaled size $k$
for a critical Galton-Watson branching process with 
a binomial offspring distribution (with a maximum of two offspring) 
and with different values of the maximum number of generations $L$.
A finite-size scaling law, of the form $D_L(k)=g(k/L^2)/L^3$, 
holds, but only for $k >10$, roughly.
Curiously, when the offspring distribution is geometric
(for which the size distribution coincides with the escape time of a random walker moving between an absorbing and a reflecting boundary \cite{Corral_garciamillan})
the deviation takes the opposite sign, see Ref. \cite{Corral_csf}.
But beyond the smallest values of $k$, 
the rescaled distributions coincide and universality shows up
\cite{Stanley_rmp}. 
}
\end{figure*}

\end{document}